\documentclass[a4paper,fleqn,usenatbib]{mnras}

\usepackage[T1]{fontenc}
\usepackage{ae,aecompl}

\usepackage{graphicx}	
\usepackage{amsmath}	
\usepackage{subfig}
\usepackage{enumitem}
\usepackage{color}
\usepackage[dvipsnames]{xcolor}

\usepackage{newtxtext,newtxmath}
\usepackage{balance}

\newcommand{\hone}{\textsc{Hi}}

\title[Synchrotron spectral index from MeerKLASS]{Measurements of the diffuse Galactic synchrotron spectral index and curvature from MeerKLASS pilot data}

\author[M. O. Irfan et al.]{Melis O. Irfan,$^{1,2}$\thanks{E-mail: mirfan@myuwc.ac.za}
Philip Bull,$^{2,1}$
Mario G. Santos,$^{1,3}$ 
Jingying Wang,$^{1}$
Keith Grainge,$^{4}$
\newauthor
Yichao Li,$^{9,1}$
Isabella P. Carucci,$^{5,6}$
Marta Spinelli,$^{7,8,1}$
Steven Cunnington$^{2}$
\\ \\
$^{1}$Department of Physics and Astronomy, University of Western Cape, Cape Town 7535, South Africa\\
$^{2}$Department of Physics and Astronomy, Queen Mary University of London, London, E1 4NS, UK
\\
$^{3}$ South African Radio Observatory (SARAO), 2 Fir Street, Observatory, Cape Town, 7925, South Africa\\
$^{4}$Jodrell Bank Centre for Astrophysics, Department of Physics and Astronomy, The University of Manchester, Manchester M13 9PL, UK
\\
$^{5}$Dipartimento di Fisica, Universit\`a degli Studi di Torino, via P.\ Giuria 1, 10125 Torino, Italy
\\
$^{6}$INFN -- Istituto Nazionale di Fisica Nucleare, Sezione di Torino, via P.\ Giuria 1, 10125 Torino, Italy
\\
$^{7}$Institute of Particle Physics and Astrophysics, Department of Physics, ETH Zurich, Switzerland
\\
$^{8}$INAF, Osservatorio Astronomico di Trieste, Via G. B. Tiepolo 11, I-34131 Trieste, Italy
\\
$^{9}$Department of Physics, College of Sciences, Northeastern University, Shenyang 110819, China
}

\date{Accepted XXX. Received YYY; in original form ZZZ}

\pubyear{2021}

\begin{document}
\label{firstpage}
\pagerange{\pageref{firstpage}--\pageref{lastpage}}
\maketitle

\begin{abstract}
21cm intensity mapping experiments are bringing an influx of high spectral resolution observational data in the $\sim100$ MHz -- $1$~GHz regime. We use pilot $971-1075$\,MHz data from MeerKAT in single-dish mode, recently used to test the calibration and data reduction scheme of the upcoming MeerKLASS survey, to probe the spectral index of diffuse synchrotron emission below 1~GHz within $145^{\circ} < \alpha < 180^{\circ}$, $-1^{\circ} < \delta < 8^{\circ}$. Through comparisons with data from the OVRO Long Wavelength Array and the Maipu and MU surveys, we find an average spectral index of $-2.75 < \beta < -2.71$ between 45 and 1055\,MHz. By fitting for spectral curvature with a spectral index of the form $\beta + c \, {\rm{ln}}(\nu / 73~{\rm MHz})$, we measure $\beta = -2.55 \pm 0.13$ and $c = -0.12 \pm 0.05$ within our target field. Our results are in good agreement (within $1\sigma$) with existing measurements from experiments such as ARCADE2 and EDGES. These results show the calibration accuracy of current data and demonstrate that MeerKLASS will also be capable of achieving a secondary science goal of probing the interstellar medium.
\end{abstract}

\begin{keywords}
Cosmology: diffuse radiation, Methods: data analysis, Radio continuum: ISM
\end{keywords}



\section{Introduction}
\label{sec:intro}

In recent years an increasing number of new radio telescopes have been coming online, a substantial fraction of which are designed to target the redshifted 21cm emission line from neutral hydrogen amongst other science goals. Due to their large sky coverage, unprecedented sensitivity and MHz frequency ranges, these telescopes provide a unique opportunity to probe the Interstellar Medium (ISM) and catalogue extragalactic radio sources across previously unexplored frequencies. The RACS survey \citep{racs} has already provided the community with $15^{\prime\prime}$ resolution images of the $-90^{\circ} < \delta < 41^{\circ}$ sky at 888\,MHz taken using the Australian Square Kilometer Array Pathfinder (ASKAP), while the GLEAM survey \citep{gleam} has mapped the $-90^{\circ} < \delta < 30^{\circ}$ sky between 76 and 222\,MHz at a resolution of $2^\prime$ using the Murchison Widefield Array (MWA). The Owens Valley Radio Observatory Long Wavelength Array \citep[OVRO-LWA;][]{lwa} provides a $15^\prime$ resolution view of the Northern sky ($\delta > -30^{\circ}$) at eight frequency snapshots between 35 and 80\,MHz, and has been used to improve the Global Sky Model of diffuse Galactic emission \citep{gsm, lwagsm}.

At frequencies below $\sim 1$\,GHz, the dominant source of diffuse Galactic emission outside of the Galactic plane is synchrotron emission \citep{cbass18}. The synchrotron emission mechanism are cosmic-ray electrons (CRe) accelerating round the Galaxy's magnetic field lines. Synchrotron emission is present in both intensity and polarisation, but we limit the focus of this study to intensity measurements only. As synchrotron emission depends on the power-law energy distribution of the CRe, the spectrum at each pixel on the sky ($p$) should also take the form of a power-law,
\begin{equation}
T_\text{sy}(\nu, p) \propto \left(\frac{\nu}{\nu_{0}} \right)^{\beta_\text{sy}(p)},
\end{equation}
where the synchrotron spectral index $\beta_{\rm sy}$, measured between the two frequencies $\nu$ and $\nu_0$, varies across the sky due to energy losses as the CRe propagate anisotropically through the Galaxy. The synchrotron spectral index also changes over frequency, a process usually referred to as spectral index ``curvature'', the largest degree of which is believed to occur under 5\,GHz \citep{e2013, plancklow}. Significant advancements towards absolutely calibrated measurements of the radio sky were first made in the late sixties and early seventies with surveys such as the 178\,MHz Northern-sky map of \cite{Turt178}, the 404\,MHz Northern-sky map of \cite{Toth404} and the 150\,MHz all-sky map of \cite{Land150}. Currently, the most frequently utilised proxy for synchrotron emission is the 408\,MHz map of \cite{oldhas} (calibrated using the \cite{Toth404} data); favoured due to its full sky coverage, detailed angular resolution of $56^\prime$ and observation at a frequency where synchrotron emission is believed to be the dominant emission across the majority of the sky. In this work we examine the $145^{\circ} < \alpha < 180^{\circ}$, $-1^{\circ} < \delta < 8^{\circ}$ region, an area of the sky close to, but outside of, the large-scale synchrotron emission feature know as Loop 1, which contains the North Polar Spur \citep{berk, nps, clivespur}. Our region itself is not associated with any distinct foreground emission, synchrotron or otherwise.

We observe that the surge in interest around 21\,cm intensity mapping is resulting in a variety of new secondary science results concerning the ISM. As next-generation facilities such as the Square Kilometer Array Observatory become operational, such insights will become commonplace. 21\,cm intensity mapping experiments measure the forbidden spin-flip transition of neutral hydrogen across redshift to probe the formation and evolution of cosmic large scale structure and the intergalactic medium \citep{batlss, chang}. These experiments measure the 21\,cm temperature fluctuations around the mean signal, either in interferometric mode -- e.g., OVRO-LWA \citep{ovro21}, Tianlai \citep{tianlai}, the Packed Ultra-wideband Mapping Array \citep[PUMA;][]{puma}, the Hydrogen Intensity and Real-time Analysis eXperiment \citep[HIRAX;][]{hirax} and the Canadian Hydrogen Intensity Mapping Experiment \citep[CHIME;][]{chime} -- or in single-dish mode -- e.g., Green Bank Telescope \citep[GBT;][]{chang2010}, Parkes \citep{anderson2018}, BAO from Integrated Neutral Gas Observations \citep[BINGO;][]{bingo} and the Five-Hundred-Meter Aperture Spherical Radio Telescope \citep[FAST;][]{fast}. See \citet{Bull:2014rha} for an overview of other recent/proposed experiments.

Experiments with the primary goal of measuring the absolute (mean) global 21\,cm temperature have also proved to be a particularly powerful resource for ISM physics. Such radiometers are absolutely calibrated, not just in terms of their temperature scale, but also with regards to their survey zero-levels. \citet{leda} determine a diffuse Galactic emission spectral index between $-2.56$ and $-2.50$ across 50 to 87\,MHz using the Large-aperture Experiment to Detect the Dark Age (LEDA); see references therein for a detailed overview of single dipole antenna measurements of the diffuse Galactic spectral index. As a direct result of these `secondary science results', \citet{padovani} were able to use the diffuse Galactic spectral index measurements from the EDGES global 21\,cm experiment \citep{edgescurve} to inform their modelling of CRe propagating through the Galactic magnetic field. Despite not being explicitly designed to probe the ISM, 21\,cm experiments unavoidably generate maps and spectra of the diffuse Galactic emission too, and in the process are also contributing to an understanding of emission mechanisms within our own Galaxy. This understanding can further contribute to systematics mitigation in other fields, for example polarised cosmic microwave background mapping \citep[e.g.,][]{Dickinson:2016xyz, Basu:2019xrm}.

In this work we use data from a recent pilot survey designed to test the calibration and observation strategy proposed for the MeerKAT Large Area Synoptic Survey \citep[MeerKLASS, ][]{calib}. The survey design and science goals for MeerKLASS can be found in \citet{meerklass}; here we simply focus on extracting information about the synchrotron spectral index from the available time-ordered data (TOD) from the pilot survey. The data used in this work are within the $145^{\circ} < \alpha < 180^{\circ}$, $-1^{\circ} < \delta < 8^{\circ}$ observation field and the $971-1075$\,MHz frequency range. 

This paper is organised as follows. In Section~\ref{sec:data} we detail the MeerKAT data used. Section~\ref{sec:method} describes the techniques used to verify the data and obtain our results. Section~\ref{sec:res} shows the spectral index and curvature values obtained from a combination of MeerKAT and ancillary data. In Section~\ref{sec:PCA}, we compare our results with those obtained from the first component of a Principal Component Analysis. In Section~\ref{sec:conc} we conclude. Throughout this paper we shall refer to \citet{calib} as `W21' for brevity.

\section{Data}
\label{sec:data}

As MeerKLASS is a \hone \, intensity mapping experiment aiming to measure 21\,cm fluctuations around the global 21\,cm average temperature, the data reduction pipeline used to calibrate the survey focuses on the stability of the gain calibration over time. Obtaining the average Galactic plus extragalactic sky temperature at each frequency is not necessary for this purpose, therefore MeerKLASS is not absolutely calibrated in terms of the zero-level of the foreground emission at each frequency. The MeerKLASS pilot survey makes use of the 64 dishes within the MeerKAT array, each of which has both HH and VV (`horizontal and vertical') linearly-polarised radio receivers. The two linear polarisation receivers from each dish measure the Stokes parameters $I + \tilde{Q}$ and $I - \tilde{Q}$, respectively, where $\tilde{Q}$ is Stokes Q as measured in the telescope reference frame. In this work we investigate Stokes I only, and so use the combined HH and VV measurements. This set-up provides 64 `independent' views of the same region of sky for each observation block when the array is used in single-dish mode. Each receiver, however, has its own unique receiver temperature which changes over frequency. We make use of the MeerKAT $971-1075$\,MHz survey TOD, calibrated into kelvin using the scheme described in W21. The MeerKLASS data reduction pipeline calibrates the data into kelvin through the use of a Bayesian parametric fit to the data, which requires priors for the receiver temperature, the diffuse Galactic emission, and any elevation-dependent emission. 

The prior for diffuse Galactic emission is provided by the Python Sky Model \citep[PySM;][]{pysm}. The impact of the choice of Galactic model on the calibration from receiver units into kelvin is minimal; for a single frequency channel, a 30$\%$ change to PySM results in less than a 2$\%$ change to the overall level of the system temperature in kelvin. In this paper, we use both the `Temperature-Temperature plot' method \citep[e.g.,][]{turtle, rod, plat, reich2004}, and spectral energy distribution (SED) models to probe the diffuse Galactic synchrotron spectral index from the MeerKLASS pilot data and a selection of ancillary data, none of which have absolutely calibrated zero-levels. 

\begin{figure}
 \centering
  {\includegraphics[width=0.98\linewidth]{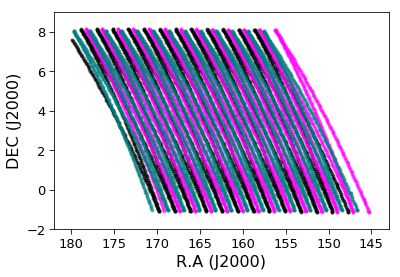}}\\
 \caption{The right ascension and declination scan patterns of the three observation blocks (different colours) used in this analysis, with coordinates in degrees.}
 \label{fig:radec}
   \end{figure} 

\begin{figure}
 \centering
  {\includegraphics[width=0.98\linewidth]{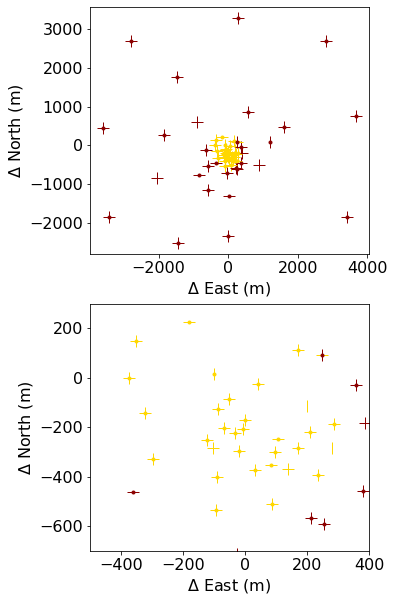}}\\
 \caption{MeerKAT dishes used for the full data set (combination of both subsets) and subsets plotted according to their East and North displacement from the telescope reference position. The gold points indicate Subset 1 and the dark red Subset 2. Dishes in use on 2019-02-25 are indicated with a horizontal line, dishes in use on 2019-04-01 are indicated with a vertical line and dishes in use on 2019-04-23 are indicated with a dot. The bottom figure presents a magnification of the compact central region within the top figure.}
 \label{fig:dnums}
   \end{figure}

In total there are seven observation blocks of  MeerKLASS pilot survey TOD available for use. As synchrotron emission is the brightest component of the total observation data, we can be selective about the data we choose to include in our analysis, i.e., we can apply relatively aggressive cuts on data quality. Out of the seven available observation blocks listed in W21, we exclude two on the basis that they are rising scans that are found to be more susceptible to radio frequency interference (RFI) within their VV polarisation measurements (see W21); one on the basis that it begins before sunset; and and one on the basis that its volume of data is already heavily reduced by the pipeline RFI flagger.\footnote{The remaining three observation blocks used in this analysis are 2019-02-25 00:40:11, 2019-04-01 22:06:17 and 2019-04-23 20:41:56; see W21 for further details.} Each of these blocks traces a slightly different coverage map across our full observation patch. The three coverage patterns are shown in \autoref{fig:radec}; the total area observed covers $35 \times 9$ square degrees at resolutions between $1.32^{\circ}$ (at 1075\,MHz) and $1.48^{\circ}$ (at 971\,MHz), due to the frequency-dependant beam. \autoref{fig:dnums} shows the available dishes that were used in the full dataset, and also defines two `jackknife' subsets that will be used in Section~\ref{sec:method} to check data consistency. The dishes are plotted according to their East and North displacements from the MeerKAT reference position of (lon, lat) = ($21^{\circ}27^{\prime}, -30^{\circ}43^{\prime}$).

RFI is a severe problem for MeerKAT auto-correlation measurements around declinations of $\delta \approx 0^\circ$, and the high-frequency channels are more susceptible to RFI contamination than the lower-frequency channels. We therefore restrict our analysis to between 971.2 and 1075.5\,MHz, with a channel width of about 0.2\,MHz.

\section{Method}
\label{sec:method}  

\subsection{The total system temperature}
\label{sec:theory}    

The total system temperature for each receiver is a combination of the receiver temperature itself ($T_{\rm  rec}$), the elevation-dependent atmospheric emission plus ground spillover ($T_{\rm  el}$), the CMB monopole ($T_{\rm CMB}$), Galactic emission ($T_{\rm G}$), and the combined extragalactic emission ($T_{\rm Ex}$), such that
\begin{equation}
\label{eq:tsys}
T_{\rm{sys}}(p, \nu) = T_{\rm CMB} + T_{\rm G}(p, \nu) + T_{\rm Ex}(p, \nu) + T_{\rm el}(\nu) + T_{\rm rec} (\nu).
\end{equation}
Telescope contributions to the system temperature such as dish surface inaccuracies and skin depth are also folded into the components above. Dish surface inaccuracies would impact the aperture efficiency and would increase the ground spillover, included above as $T_{\rm  el}$. The receiver skin depth, the depth within a conducting surface through which electromagnetic radiation can penetrate, is frequency-dependant and impacts $T_{\rm  rec}$ (although, negligibly at $\sim 1$\,GHz for MeerKAT). We assume that the CMB anisotropies are negligible in comparison with the Galactic synchrotron emission temperature \citep{santos05} and so we take the value of $T_{\rm CMB}$ to be a constant 2.725\,K. We also ignore the 21\,cm contribution for the same reason. While the Galactic and extragalactic temperatures change across both pixel ($p$) and frequency ($\nu$), the receiver temperature and elevation-based contributions are expected to have a substantially greater spatial stability (W21). 

At $\sim 1$\,GHz frequencies and high Galactic latitudes, the dominant diffuse emission is expected to be diffuse synchrotron emission. This assumption was verified using the ratio of predicted free-free emission to predicted synchrotron emission for the MeerKLASS pilot survey patch at 1075\,MHz. The predicted emission amplitudes were calculated using the Planck Legacy Archive\footnotemark \footnotetext{\url{http://pla.esac.esa.int/pla}} FFP10 simulations of free-free and synchrotron emission at 30\,GHz, smoothed to a 1$^{\circ}$ FWHM resolution and scaled to 1075\,MHz using spectral indices of $-2.1$ and $-2.9$, respectively. The predicted free-free emission was on average less than 1$\%$ of the synchrotron emission temperature. When using the predicted foreground maps to recover the $-2.9$ spectral index, the inclusion of free-free emission was found to impact the recovered spectral index by $0.01-0.025$ over the 970 to 1075\,MHz frequency range. 

\subsection{Extragalactic contributions}
\label{sec:eg}  

The extragalactic contributions to the total system temperature largely come from point sources, with three main components. First, there is an overall constant temperature offset resulting from the mean value of the unresolved point sources; then, there is the clustering contribution from partially-resolved sources; and finally, there is a Poisson-like contribution of individually-resolved point sources \citep{batps}. The mean value of the unresolved extragalactic point sources (ueps) is a spatially constant offset value, $T_{\rm ueps}(\nu)$, and is by far the largest contribution. As such, we choose to treat the spatially varying Poisson and clustering contributions as insignificant by including them in our estimate for Galactic synchrotron emission i.e., our diffuse synchrotron emission estimate will also absorb temperature contributions from partially-resolved point sources,
\begin{equation}
T_{\rm G}(p, \nu) + T_{\rm Ex}(p, \nu) = T_{\rm sync}(p, \nu) + T_{\rm ueps}(\nu).
\end{equation}
This means that a key assumption in our analysis is that the Poisson and clustering point source contributions are minor contributions in comparison to the variations in diffuse Galactic synchrotron emission. Going back to Eq.~\ref{eq:tsys}, we now see that 
\begin{equation}
\label{eq:newtsys}
T_{\rm{sys}}(p, \nu) = T_{\rm CMB} + T_{\rm sync}(p, \nu) + T_{\rm ueps}(\nu) + T_{\rm el}(\nu) + T_{\rm rec} (\nu).
\end{equation}
For reference, \citet{psg} collate absolute brightness measurements of $T_{\rm ueps}(\nu)$ across both MHz and GHz frequencies, finding 0.91\,K at 610\,MHz and 0.10\,K at 1.4\,GHz using a set of three radiometers (`TRIS'). They produce a model fit to these data which predicts a 0.26\,K offset temperature at 970\,MHz due to unresolved extragalactic point sources. 

\subsection{Temperature-Temperature plots}
\label{sec:eg}  

In this section, we review the `TT-plot' method, which allows the spectral dependence of the dominant synchrotron emission component to be separated out from other spatially/spectrally-constant contributions to multiple datasets, even in the absence of an absolute zero-point calibration for each dataset.

As a first approximation, diffuse Galactic synchrotron emission can be characterised by a power law,
\begin{equation}\label{eq:synch}
T_{\rm{sync}}(p, \nu_{2}) = T_{\rm{sync}}(p, \nu_{1}) \times m(\nu_1, \nu_2; \beta_{\rm sync}),
\end{equation}
where
\begin{equation}
m = \left(\frac{\nu_{2}}{\nu_{1}} \right )^{\beta_{\rm sync}}.
\label{eq:defm}
\end{equation}
For two surveys, both with absolutely-calibrated zeros-levels, conducted at frequencies $\nu_{1}$ and $\nu_{2}$ respectively, the emission spectral index can simply be calculated at each pixel as
\begin{equation}
\beta(p) = \frac{{\rm{ln}}(T_{\rm{sync}}(p, \nu_{2}) / T_{\rm{sync}}(p, \nu_{1}))}{\rm{ln}(\nu_{2} / \nu_{1})}.
\label{eq:perpix}
\end{equation}
However if the zero-level is {\it not} determined by the minimum synchrotron emission temperature at that frequency, and is instead, for instance, dominated by several kelvin of receiver temperature for one survey, and a different level of receiver temperature for the other, then \autoref{eq:perpix} is of no use and the spectral index cannot be determined for a single pixel on the sky. The two data sets can still be used to determine the emission spectral index however, as long as the change in temperature across several pixels, at each frequency, is solely due to Galactic synchrotron emission. This is because the temperature distribution across pixels also follows a power law:
\begin{align}
\Delta T_{\rm{sync}}(\nu_{2}) &= T_{\rm{sync}}(p_2, \nu_{1}) \left(\frac{\nu_{2}}{\nu_{1}} \right )^{\beta + \delta_{2}} - T_{\rm{sync}}(p_1, \nu_{1}) \left(\frac{\nu_{2}}{\nu_{1}} \right )^{\beta + \delta_{1}}, \nonumber\\ 
&\approx \Delta T_{\rm{sync}}(\nu_{1}) \left(\frac{\nu_{2}}{\nu_{1}} \right )^{\beta},
\label{eq:tiny}
\end{align}
where $\Delta T_{\rm{sync}}$ is the temperature difference between pixel $p_1$ and $p_2$, and $\delta_{1}$ and $\delta_{2}$ represent the small deviations from the mean spectral index. As long as $\delta_{1}$ and $\delta_{2}$ are small, they can be considered negligible, and the approximation in \autoref{eq:tiny} holds.    

\citet{wehus} demonstrated that if the total measured system temperatures are not entirely dominated by synchrotron emission, but instead consist of synchrotron emission plus additional spatially-constant offset contributions, then the linear regression of observations at different frequencies will separate the spectral index of the spatially-varying emission from
the spatially constant offsets at each frequency,
\begin{equation}
T_{\rm{sys}}(\nu_{2}) = m \times T_{\rm{sys}}(\nu_{1}) - m \times T_{\rm off1} + T_{\rm off2}.
\label{eq:weeq}
\end{equation}
The gradient of the linear fit is a function of the emission spectral index only while the y-intercept is dependant on both the spectral index and the spatially constant offsets. The T-T plot method requires the data under analysis to be scaled with a single spectral index value, therefore this method must be restricted to spatial regions small enough to ensure limited variations of the synchrotron spectral index. T-T analysis can be seen to work perfectly in the case of noiseless simulation data, however for empirical data the inclusion of instrumental noise, even if only Gaussian, introduces degeneracies between the fitted gradient and y-intercept. The error on the spectral index is calculated from \autoref{eq:defm} by propagating the error on the fitted gradient $m$, such that
\begin{equation}
 \label{eq:erreq}
\sigma_{\beta} = \frac{\sigma_{m}}{m \, \rm{ln} \left(\nu_{2} / \nu_{1} \right )}.
\end{equation}
While noise-induced errors on the fitted gradient increase the uncertainty on the spectral index, the larger the difference between $\nu_2$ and $\nu_1$ the smaller the impact on the inferred spectral index.

\subsection{Adapting the T-T plot method}
\label{sec:pilot}    

A verification of the individual receiver temperatures was presented in W21 using the T-T plot method. For MeerKAT, the receiver temperatures vary somewhat smoothly over the full frequency range, forcing us to select $\nu_{1}$ and $\nu_{2}$ values that are less than a few MHz apart, where the receiver temperature should be close to constant. As the typical MeerKAT receiver temperatures are an order of magnitude larger than the diffuse synchrotron emission present ($\sim 6 - 10$\,K, as opposed to $0.3 - 0.9$\,K), we were able to verify the MeerKLASS calibration pipeline estimates of the receiver temperatures to within 1$\%$ ($\sim 0.1$\,K) using the T-T plot method (shown in figure 20 of W21). To study diffuse synchrotron emission specifically however, either greater precision on the linear regression gradients is required, or a larger $\nu_{2}$/$\nu_{1}$ ratio is needed. As this larger ratio cannot be provided by the MeerKLASS data alone, a simple solution is to bring in an additional data set.

The 408\,MHz all-sky map of \cite{oldhas} provides a $56^\prime$ resolution view of synchrotron emission across the full sky. We use the reprocessed version of this map \citep[destriped and desourced;][]{newhas} to extend our observational frequency range. We choose to use the reprocessed map as it addresses the main concerns of the original data; namely, the resolution and map-making artefacts. The original 408\,MHz data were cited with a resolution of $51^\prime$, however, as the data were a composite of four separate partial-sky surveys there was some ambiguity around this number, which has since been corrected to  $56 \pm 0.6^\prime$ by \cite{newhas}. The original data were also subject to spurious artefacts, visible by eye in the full-sky map and following the scan direction, caused by $1/f$ noise. \cite{newhas} reprocessed the 1080 $\times$ 540 ECP map of the MPIfR Bonn survey sampler \footnote{\url{http://www.mpifr-bonn.mpg.de/survey.html}} and reduced the amplitude of these artefacts by more than $70\%$. Both the original and reprocessed data are cited with a $10\%$ absolute calibration uncertainty. For the purposes of our analysis the most damaging systematic would be the spurious artefacts in the data; although the calibration uncertainty is large it is a constant factor. Map-making artefacts, on the other hand, introduce spurious structure which could be mistaken as synchrotron emission. The amplitude of these artefacts was measured to be $\sim 1$\,K at 408\,MHz \citep{rod}, which scales to 0.096\,K at 970\,MHz if using a spectral index of -2.7. As we expect to measure synchrotron emission between 0.3 and 0.9\,K across our frequency range, a $\sim 0.1$\,K artefact would be a notable contaminant. By using the reprocessed map, the level of these artefacts were significantly reduced to the extent that they have no noticeable impact on the results presented in the sections to come.

To use the \cite{newhas} data for our purposes, we extract a 2D array of temperatures from the full sky map which corresponds to our observational area by using standard HEALPix projection functions \citep{healpix}. We then mask out regions of the map within $0.3^\circ$ of known radio point sources with flux densities greater than 1\,Jy at 1.4\,GHz, leaving just the diffuse emission pixels. For each T-T plot, only a subset of these diffuse emission pixels will be used, as there needs to be the equivalent MeerKLASS pilot data observed in each pixel, and this depends on the scan strategy and the observational block being used. The MeerKLASS pilot data also have different angular resolutions in each frequency channel due to the frequency-dependant beam. We fit a 2D Gaussian approximation to the true beam pattern, giving an approximate Gaussian FWHM of $1.32^\circ$ to $1.48^\circ$ for the frequencies between 971 and 1076\,MHz. We then smooth the Haslam data to the lowest common resolution of $1.48^\circ$ for this analysis.

      \begin{figure}
      \centering
            {\includegraphics[width=0.92\linewidth]{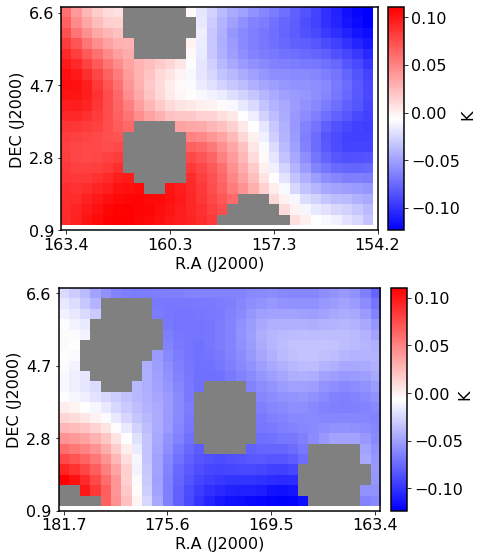}}\\
      \caption{GSM model of total diffuse emission at 971\,MHz and a resolution of $56^\prime$ within our observation patch. The patch has been split into two (note the RA ranges) and each sub-region has been mean-centred. The missing circular regions have been excluded by our point source mask.}
         \label{fig:gsm}
   \end{figure}

\begin{figure*}
\centering
{\includegraphics[width=0.97\linewidth]{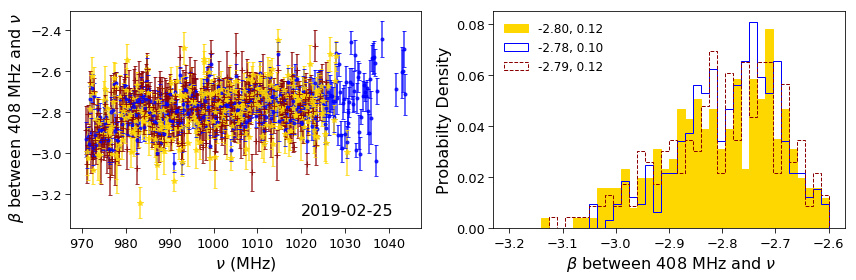}}\\
{\includegraphics[width=0.97\linewidth]{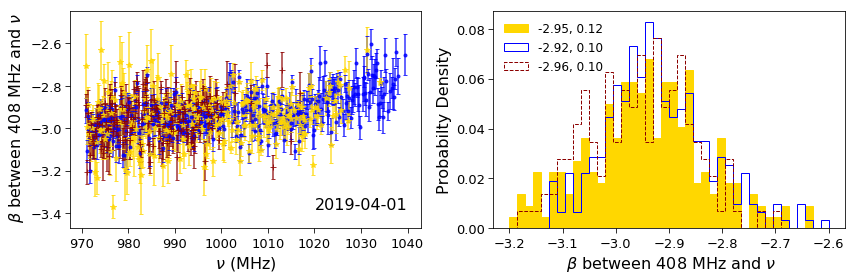}}\\
{\includegraphics[width=0.97\linewidth]{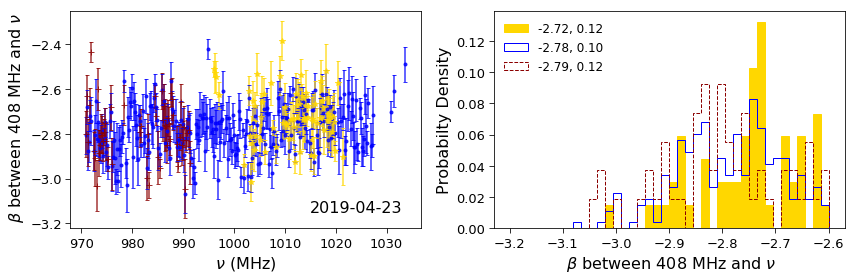}}\\
 \caption{{\it{Left:}} The $408-\nu$\,MHz mean spectral indices across all dishes. The error bars are the standard errors on the mean in each channel. {\it{Right:}} The histogram of the mean spectral indices across frequency. The legends display the mean spectral index and the standard deviation of these indices across all frequencies. In all plots the blue data points are for the full set of dishes, the gold are for Subset 1, and the dark red for Subset 2.}
         \label{fig:jacks}
   \end{figure*} 

As we intend to use linear regression, there is a requirement for the MeerKAT channels to be strongly correlated with the reprocessed Haslam data. A comparison between each data channel (for each MeerKAT dish) and the Haslam data was made, and if the correlation coefficient between the two was less than 0.85 we took this as an indicator of residual RFI contamination (or other systematic contamination), causing us to reject that frequency channel of that dish during that observation block from the rest of the analysis.

When constructing the T-T plots used in this section (and in Section~\ref{sec:TT}), we also wanted to select a smaller region within our total observation patch so as to minimise the spatial variations of the spectral index encountered. We cut our patch in half (in terms of RA) and selected the lower patch ($154^{\circ} < \alpha < 163^{\circ}$) for all our T-T plot analyses. \autoref{fig:gsm} explains the motivation for this cut -- it shows the total emission model for our observational patch at 971\,MHz, at a resolution of $56^\prime$, from the Global Sky Model \citep{gsm}. The patch has been split into two, and each sub-region has been mean-centred. The data within the $154^{\circ} < \alpha < 163^{\circ}$ region shows more spatial structure, and so will provide a larger range of temperatures for our T-T plots, allowing us to gain a better handle on the gradient. We also show examples of T-T plots taken within the $163^{\circ} < \alpha < 182^{\circ}$ region in Appendix~\ref{sec:apA} to demonstrate how the minimal spatial structure prevents a good linear fit to the data from being obtained.  

The MeerKAT data have an additional complication in that Galactic and extragalactic emissions are not the only spatially varying contributions to the total system temperature. W21 modelled an additional temperature component which varies smoothly in time that can be attributed to some combination of low-level RFI and ground emission. This component prevents the standard T-T plot method from being used, as the offset term is no longer constant in time. To combat this problem the data were divided into smaller chunks in time to ensure that the offset term is approximately constant within each chunk, even if it does drift across chunks. The data were divided into chunks of 30\,s and the $408-\nu$\,MHz spectral indices were calculated for each of them. For one observation block, one dish, and one frequency channel, we therefore calculate several $408-\nu$\,MHz spectral indices. A weighted mean, based on the fitted spectral index errors $\sigma_\beta$, is then used to provide the final spectral index estimate for each channel.

Note that the chosen chunk size of 30\,s was determined empirically through simulations. We detail these simulations in Appendix~\ref{sec:apB}, where we also verify that our implementation of the T-T plot method can successfully measure the synchrotron spectral index to within $\pm 0.04$. While the simulations are a simplification, in that they do not feature small-scale effects such as the frequency-dependant beam and thermal noise, they are able to demonstrate the stability of our method in the face of time-varying terrestrial temperature components of order $\sim 10\%$ of the diffuse Galactic emission temperature variations. It should also be noted that additional contaminants, which have not been modelled in these simulations, could arise over the course of the observational blocks and may be present in the real data, some examples being RFI/satellite contamination low-level enough to slip through the RFI filters, or ground pick-up with a different amplitude or time-dependant behaviour to that of our ground pick-up model. In the next section we move beyond the simulated data to apply the time chunk-based T-T plot method to the observational MeerKLASS pilot data.

\subsection{Jackknives}
\label{sec:jacks} 

Each MeerKAT data channel can be used alongside Haslam data to provide a spectral index measurement at each channel, for each dish and for each observation block. These measurements can be used to check data consistency over frequency, since the spectral index shouldn't change noticeably across the MeerKAT sub-band and across the different observation blocks. We can also check that our 408-1075\,MHz synchrotron spectral indices are consistent with the literature on synchrotron radiation within the ISM. The synchrotron spectral index depends on the slope of the CRe energy distribution ($s$) like so: $\beta = (s - 3)/2$ and, for typical Galactic magnetic field strengths ($2-20\,\mu$G), $s$ can range between $-1.94$ and $-3.23$ giving an expected spectral index range of $-3.4 < \beta < -2.5$ \citep{e2018, padovani}. \autoref{fig:jacks} plots the $408-\nu$\,MHz spectral index as a function of frequency for the three observation blocks (top to bottom), and for the full set and both subsets of dishes (see Fig.~\ref{fig:dnums}). Histograms of the spectral index distributions are shown next to each frequency plot. For each channel, the $408-\nu$\,MHz spectral index plotted is the average value from all the dishes used, as long as a minimum of ten dishes were left after cuts. It can be seen that the subsets do not have enough dishes to provide spectral index estimates at every channel. Anomalous indices were identified as those $> 2.7 \times {\rm MAD}$ (the median absolute deviation) away from the median value, and were left out of the averaging. Using the MAD of a data set to separate signal from noise is a common technique in wavelet-based component separation, e.g., \citet{threshold}; MAD-based thresholding can typically be used to identify outliers at between $2.5$ and $3 \times$ the MAD above/below the median value \citep{kmad}.

\begin{figure}
      \centering
            {\includegraphics[width=0.9\linewidth]{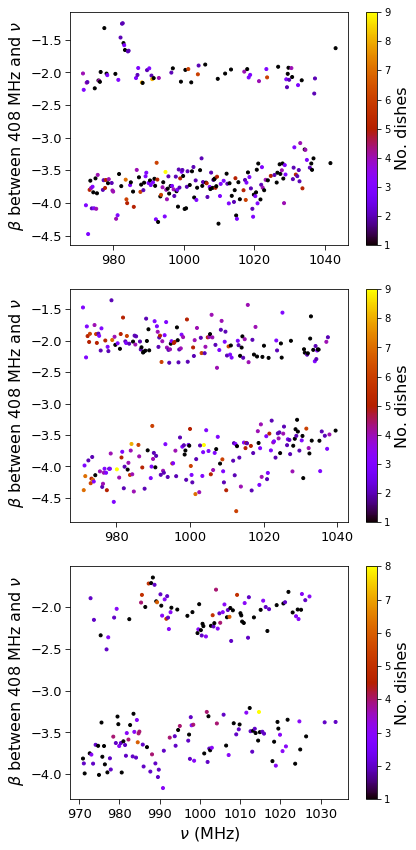}}\\
      \caption{Anomalous measurements of the synchrotron spectral index excluded from the averaging used to produce \autoref{fig:jacks}. The most extreme values are plotted at each frequency and the colour scale indicates the number of dishes which produced anomalous spectra index measurements.}
         \label{fig:odds}
   \end{figure} 
The most extreme of the anomalous indices excluded from the mean calculation are show in \autoref{fig:odds}, where we plot the spectral index furthest from the mean value as a function of frequency. While the most extreme anomalous index at each frequency is only calculated using measurements from a single dish, more than one dish may present anomalous indices. The colour scale on the scatter plots shown in \autoref{fig:odds} represents the number of dishes for which an anomalous spectral index was calculated.
   
In Appendix~\ref{sec:apB}, we saw that for simulation data the synchrotron spectral index could be measured to an accuracy of 0.04 for a single MeerKAT dish when the main data contaminant was a smoothly time-varying component of magnitude $\sim 10\%$ of the synchrotron temperature variations. Both \autoref{fig:jacks} and \autoref{fig:odds} reveal that the empirical data contain temperature components not accounted for within our simulation test-bed. Some dishes, up to a maximum of nine, can be seen in \autoref{fig:odds} to present spectral indices steeper than $-3.5$ or shallower than $-2.5$. The fact that there is no discernible pattern between dish geographical location and the occurrence of anomalous spectral index measurements seems to hint toward the existence of low-level RFI within the data rather than unmodelled ground pick-up. No clear understanding of this spurious temperature contribution has yet been reached however. Similarly, in \autoref{fig:jacks}, at any given frequency the error on the mean spectral index, as calculated from all the contributing dishes, is larger than what we would expect from simulations alone.

Nevertheless, we can use these jackknives to demonstrate that no significant systematic biases exist across the MeerKAT dishes. \autoref{fig:jacks} demonstrates that no difference is seen in the spectral index measurements of either of our two dish subsets when compared to the results obtained from the full complement of available dishes. The first and third rows of \autoref{fig:jacks} indicate a mean spectral index between 408\,MHz and 1\,GHz of $-2.80 < \beta < -2.72$. While the second row shows that one of the observation blocks favours a slightly steeper spectral index of $\sim -2.95$ for the same frequency range. This $\sim 5\%$ discrepancy between the three observation blocks is comparable to the $\sim 5\%$ variation of the spectral index across frequency seen for each observation block and each dish set and subset. Having shown the pilot data to be consistent at the $5\%$ level, comparisons can now also be made with ancillary data sets.

\subsection{Map making}
\label{sec:mm}

In the previous sections we outlined the rationale behind the T-T plot method and applied it to individual TOD. In this section we mean-centre and combine the TOD to produce a cube of 2D spatial variations across frequency. Rephrasing \autoref{eq:newtsys} to explicitly consider each temperature component as a mean value plus fluctuations around this mean gives
 \begin{align}
 \begin{split}
T_{\rm{sys}}(p, \nu) &= \delta T_{\rm sync}(p, \nu) + \overline{T}_{\rm sync}(\nu) + \delta T_{\rm terr}(p, \nu) + \overline{T}_{\rm terr}(\nu) \\ &+ \overline{T}_{\rm ueps}(\nu) + T_{\rm{cmb}},
\end{split}
 \end{align}
where the CMB anisotropies are considered negligible and the receiver and elevation-dependant temperature contributions are represented jointly as the terrestrial contributions, $T_{\rm{terr}}$. As discussed in Section~\ref{sec:eg}, we make no attempt to separate the point source temperature deviations from the larger diffuse synchrotron emission temperature deviations, and consider both contributions to be represented by the term $\delta T_{\rm sync}(p, \nu)$. Mean-centering the data at each frequency, we obtain
$\Delta T(p, \nu) = T_{\rm{sys}}(p, \nu) - \overline{T}_{\rm{sys}}(\nu)$,
where
$\overline{T}_{\rm{sys}}(\nu) = \overline{T}_{\rm{sync}}(\nu) + \overline{T}_{\rm{terr}}(\nu) + \overline{T}_{\rm{ueps}}(\nu) + T_{\rm{cmb}}$.
This leaves the residual temperature in each pixel to represent the combined synchrotron and terrestrial temperature deviations. For two dishes observing during the same observation block, both the synchrotron temperature mean and deviation are in principle identical, as both dishes are tracking the same RA and declination, so that
\begin{align*}
&\Delta T_{\rm{dish1}}({\rm obs \, block \, 1}) = \delta T_{\rm sync}(p, \nu) + \delta T_{\rm terr1}(p, \nu), \\
&\Delta T_{\rm{dish2}}({\rm obs \, block \, 1}) = \delta T_{\rm sync}(p, \nu) + \delta T_{\rm terr2}(p, \nu). 
\end{align*}
Stacking the dishes from a single observation block averages out the receiver temperature variations (which should be uncorrelated) while preserving the synchrotron fluctuations on the sky.

To combine the measurements from the three observation blocks we follow the approach of W21, where pixel measurements within 0.3 degrees of each other were averaged together. From  \autoref{fig:radec} it can be seen that the similar coverage patterns of all three observations will ensure that $\overline{T}_{\rm{sync}}(\nu)$ is the same for all three observation blocks. In this case, even though the synchrotron temperature fluctuations are all slightly different, all three represent fluctuations around the same mean value and so can be averaged together within a suitable fraction of the beam width.

We summarise our method to produce a map of diffuse synchrotron emission fluctuations around the mean synchrotron temperature at each frequency as follows:
\begin{enumerate}[label*=\arabic*.,leftmargin=0.5\parindent]
  \item For each observation block, each MeerKAT dish, and each frequency channel: Subtract the mean map value to make a temperature fluctuation map.
  \item For each observation block: Average the temperature fluctuation maps from the $\sim 64$ dishes at each frequency and pixel. (Anomalous values are identified as those $2.7 \times$ the MAD above/below the mean value and are left out of the averaging.)
  \item Combine the observational blocks by averaging together all observations within 0.3 degrees.
\end{enumerate}
The resulting 3D data cube provides an estimate of the diffuse synchrotron emission temperature fluctuation that is centred at zero for each frequency. The two spatial dimensions cover the 145$^{\circ} < \alpha < 180^{\circ}$, $-1.5^{\circ} < \delta < 8^{\circ}$ region at a resolution of 0.3 degrees per pixel, and the third dimension charts the change in temperature fluctuations across the 971.2 -- 1075.5\,MHz frequency range.
 
\section{Results}
 \label{sec:res}
We now look to compare the synchrotron temperature fluctuation maps from the MeerKLASS pilot survey with those of surveys other than Haslam, using both a T-T plot analysis and spectral energy distribution (SED) models. For a proper analysis of the synchrotron spectral index in pixel space, we require all frequency maps to be at the same resolution. A subset of the total observation area is selected for smoothing to avoid empty pixels near the edges of the scan pattern. The $154^{\circ} < \alpha < 173^{\circ}$, $0.6^{\circ} < \delta < 6.5^{\circ}$ region of the above map is selected and then smoothed with a 2D Gaussian to provide a MeerKAT data cube at both 1.8$^{\circ}$ and 5$^{\circ}$ resolution. The smoothing takes the frequency-dependant beam into account, but uses an approximate Gaussian FWHM to account for the angular resolution in each channel. These resolutions were chosen to match the resolutions of the ancillary data presented in the next section.

As not all of the available MeerKLASS pilot data were used in this analysis (only 3 out of the 7 observation blocks were included), some pixels in our 3D data cube contain no information. These pixels required inpainting to enable smoothing to a common resolution, which was done using the average temperature from neighbouring pixels. The inpainted pixels were then masked out again to ensure only measured data contributed to the analysis. 

    \begin{figure}
      \centering
            {\includegraphics[width=0.99\linewidth]{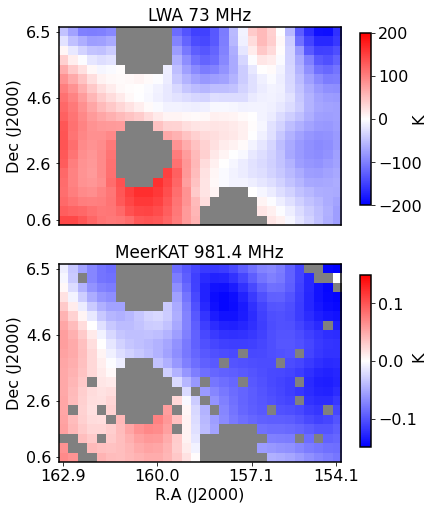}}\\
      \caption{1.8$^{\circ}$ resolution images of our observation patch as seen at 73 and 981 MHz. The data are in kelvin and have been mean-centred. The missing circular regions have been excluded by our point source mask.}
         \label{fig:ims}
   \end{figure}

\subsection{Ancillary data}
\label{sec:anc}

For our T-T plot analysis, we rely on two ancillary data maps -- one at 73\,MHz from the OVRO Long Wavelength Array (OVRO-LWA), and one at 45\,MHz from the Maipu/MU surveys. Neither of these surveys has an absolutely calibrated zero-level, which is why negative temperature values can be seen for the LWA data for example. The OVRO-LWA is a drift-scan interferometer mapping the sky above declination $-30^{\circ}$, between 35 and 80\,MHz and with a $15^\prime$ angular resolution \citep{lwa}. The LWA suffers the same blight as MeerKAT; to combat RFI, they remove spherical harmonics with $m=0$ and $m=1$, $\ell > 100$ from their data. \citet{guz} combine two 45\,MHz Northern \citep[MU;][]{45north} and Southern \citep[Maipu;][]{45south}  hemisphere surveys to present an almost full sky (96\%) map of Galactic emission at a resolution of 5$^{\circ}$. Both the Maipu and MU surveys were mapped using antenna arrays and the data processing for both took place at the Maipu Radio Observatory, Chile. The combined Maipu/MU map was calibrated using a strip of the sky common to both the Northern and Southern surveys. 

In Section~\ref{sec:apr} we compare our results on the synchrotron spectral index curvature to curvature values determined using the ARCADE2 \citep{kogut} and EDGES \citep{edgescurve} experiments. The Absolute Radiometer for Cosmology, Astrophysics and Diffuse Emission 2 (ARCADE2) was a balloon-borne experiment with five observation channels: 3.3, 8.3, 10.2, 30 and 90\,GHz. ARCADE2 had an internal, full-aperture, blackbody reference load, thus enabling the survey to have an absolutely calibrated zero-level \citep{arcade}. The Experiment to Detect the Global EoR Signature \citep[EDGES;][]{edges1} is a dipole antenna observing within both the $50-100$\,MHz and the $90-190$\,MHz bands. In order to measure the global Epoch of Reionisation (EoR) signal the survey required an absolutely calibrated zero-level allowing the EDGES team to also contribute notably to ISM physics, an example being the use of EDGES data to recalibrate existing radio surveys \citep{abscaledge}.

\subsection{T-T plots}
\label{sec:TT}

As discussed in Section~\ref{sec:pilot}, we find the RA range between 154$^{\circ}$ and 163$^{\circ}$ optimal for our T-T plot analysis. This, added to the fact that we avoid the edges of the scan strategy, results in the following analysis region for this section: $154^{\circ} < \alpha < 163^{\circ}$, $0.6^{\circ} < \delta < 6.5^{\circ}$.

    \begin{figure}
      \centering
            {\includegraphics[width=0.85\linewidth]{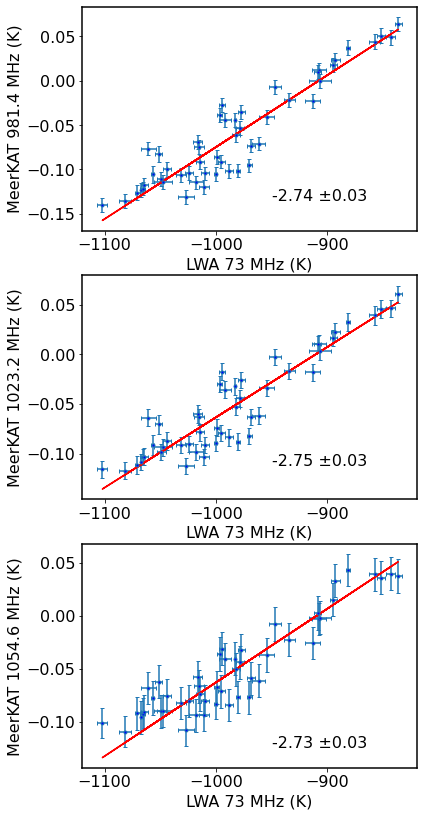}}\\
      \caption{T-T plots between MeerKLASS pilot data at 981, 1023 and 1055\,MHz and LWA data at 73\,MHz.}
         \label{fig:tts}
   \end{figure}

\autoref{fig:ims} shows our observation region at $1.8^\circ$ resolution as seen at 73\,MHz by LWA and at 981\,MHz by MeerKAT. The mean-centred temperatures are plotted, with the regions excluded from the analysis by our point source mask, formed using the NASA/IPAC Extragalactic Database,\footnote{\url{http://ned.ipac.caltech.edu}} seen as circular grey areas. \autoref{fig:tts} shows the linear regression between LWA 73\,MHz data and MeerKAT 981, 1023, and 1055\,MHz data. Both datasets have been smoothed to $1.8^\circ$ resolution with a Gaussian kernel. To avoid correlated errors, we group the pixels into bins of $3^{2}$ and plot the bin averages and standard error on these means. The spectral index of each linear regression, calculated from the fitted gradient, is shown on each plot. The total error on each spectral index is calculated using \autoref{eq:erreq}, where
\begin{equation}
\delta m = \sqrt{(\delta x)^{2} + (\delta y)^{2} + (\delta f / f)^{2}},
\end{equation}
where $\delta x$ is the LWA calibration uncertainty of 5\% \citep{lwa}, $\delta y$ is the MeerKLASS calibration uncertainty of 2\%(W21), and $\delta f / f$ is the error on the fitted gradient divided by the fitted gradient value. Larger error bars can be seen on the binned MeerKAT pixels at 1054.6\,MHz than at lower frequencies because the number of missing pixels due to RFI flagging increases with frequency.

       \begin{figure}
      \centering
            {\includegraphics[width=0.8\linewidth]{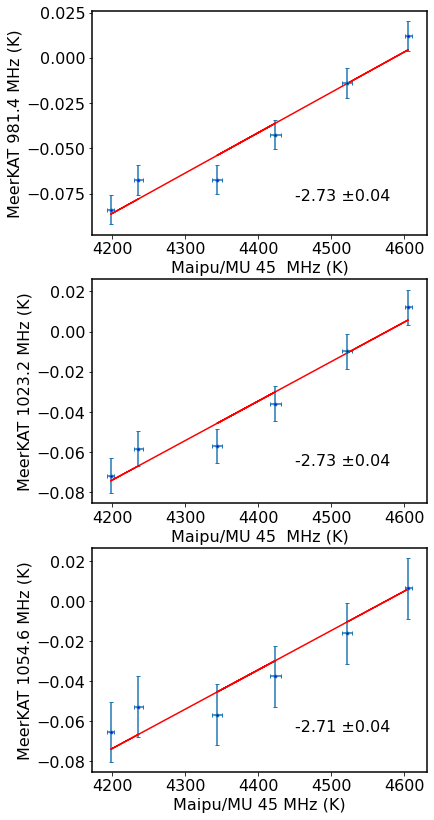}}\\
      \caption{T-T plots between MeerKLASS pilot data at 981, 1023, and 1055\,MHz and Maipu/MU data at 45\,MHz.}
         \label{fig:tts5}
   \end{figure}

\autoref{fig:tts5} shows the linear regression between Maipu/MU 45\,MHz data and MeerKAT 981, 1023, and 1055\,MHz data. In this case the MeerKLASS pilot data have been smoothed to $5^\circ$ resolution and the data have again been grouped into bins of $9^{2}$ pixels to reduce correlated errors. Ideally an even larger number of bins would have been used in this case, but our pilot observation patch is not large enough to allow this. Therefore, the data points shown in the three plots of \autoref{fig:tts5} are still partially correlated. Comparison of each of the plots in both \autoref{fig:tts} and \autoref{fig:tts5} reveals $\sim 1\sigma$ agreement in all cases for a spectral index of $-2.75 < \beta < -2.71$ between 45 and 1055\,MHz.   
   
          \begin{figure}
      \centering
            {\includegraphics[width=0.99\linewidth]{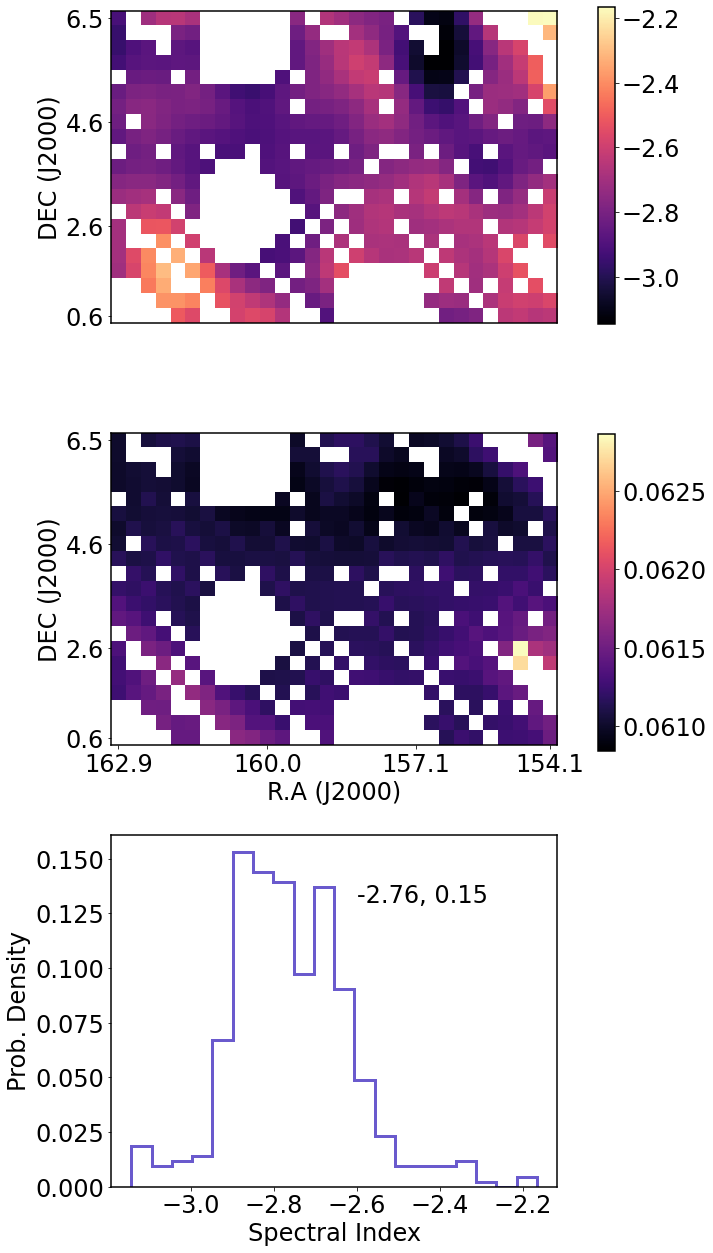}}\\
      \caption{The per-pixel spectral index between 971 and 1075\,MHz as calculated from MeerKAT data with an absolute zero-level calibration tied to the Haslam 408\,MHz monopole of $8.9 \pm 1.3$\,K. {\it{From top to bottom:}} the spectral index values, the spectral index errors, the spectral index histogram distribution. Point source masking has been applied and only pixels with more than 400 frequency measurements were allowed to contribute to the analysis. The mean spectral index and standard deviation ($-2.76$ and $0.15$ respectively) are stated on the histogram plot.}
         \label{fig:allmk}
   \end{figure}

\subsection{Bootstrapping a zero-level calibration}
\label{sec:otf}
   
Although the MeerKAT radiometers are not equipped with their own zero-level reference, like the ARCADE2 instrument has for instance \citep{arcade}, it is possible to recover the true synchrotron monopole at each frequency through the use of either multiple additional datasets or a linear regression from a survey which {\it does} have an absolutely calibrated zero-level \citep{jonas, gem}.

Through the linear regression of multiple radio frequency datasets, \citet{wehus} obtain a zero-level correction for the Haslam 408\, MHz data of $8.9 \pm 1.3$\,K. Subtracting this value from the Haslam data results in a map whose minimum temperature value represents the lowest synchrotron emission temperature at 408\,MHz. We can then calculate the MeerKAT zero-level at each frequency by performing the T-T plot linear regression with this version of the Haslam map and using \autoref{eq:weeq}, but now with $T_{\rm off1}$ equal to 0\.K. The T-T plot method fits an average, single spectral index for an entire region; we are suggesting to use the zero-levels determined through this method to make a synchrotron emission cube from solely MeerKAT data. We shall analyse the accuracy of this cube by determining the spatially varying synchrotron spectral index across the MeerKAT frequencies at each map pixel. Using the assumption of an average spectral index to produce a map of spatially-varying spectral indices will introduce an inherent bias into the method but we never-the-less find this to be a constructive exercise in evaluating our per-pixel temperature data at each frequency.  

This technique is demonstrated in \autoref{fig:allmk}, where the synchrotron spectral index between 971 and 1076\,MHz at each pixel is shown. The spectral index was calculated using \autoref{eq:perpix}, as applied to the MeerKAT data with an absolute zero-level scale tied to the Haslam 408\,MHz monopole of $8.9 \pm 1.3$\,K. Point source masking was applied, and only pixels with more than 400 frequency measurements were allowed to contribute to the analysis. This limit on the minimum number of channels was determined empirically to be high enough to prevent large spectral index errors per pixel and low enough so as not to needlessly exclude too many of the map pixels. The top panel of \autoref{fig:allmk} shows the spectral index in each pixel, the middle plot shows the error on the spectral index, and the lower panel shows a histogram of the spectral index. Good agreement can be seen between the mean spectral index of $-2.76$ between 971 and 1076\,MHz, with a standard deviation of 0.15, for the per-pixel calculation, and the spectral index range of $-2.75 < \beta < -2.71$ between 45 and 1055\,MHz from the T-T plot analysis. 

The small frequency lever-arm of 971 to 1076\,MHz makes the spectral index results very sensitive to the noise on the temperature measurements in each pixel, but the large number of channels available prevents the measurements from being noise dominated. The results in \autoref{fig:allmk} show values of the spectral index that are likely to be spurious, such as those with $\beta < -2.3$ or $\beta > -3.0$. These values do not have large ($> 0.2$) errors associated with them, confirming an actual bias in our method. The T-T plot method used to determine the MeerKAT zero-level at each frequency assumes both a single spectral index and a single offset value for the region in question. We know neither of these assumptions are strictly true and \autoref{fig:allmk} highlights regions where this technique produces the largest spectral index outliers. The spectral index map in \autoref{fig:allmk} is not shown in order to be used as-is, as a spectral index map, but rather to highlight the use of the MeerKLASS data for ISM physics when combined with ancillary data sets. When the full MeerKLASS survey is complete, the high sensitivity temperature maps and thinly spaced and numerous frequency channels can be used alongside existing survey data to extend current foreground emission knowledge into previously unmeasured frequency ranges. In addition, as we expect to move away from the $\delta \approx 0^\circ$ region, we hope to be able to use a larger percentage of the frequency channels.

   \begin{figure}
      \centering
            {\includegraphics[width=0.99\linewidth]{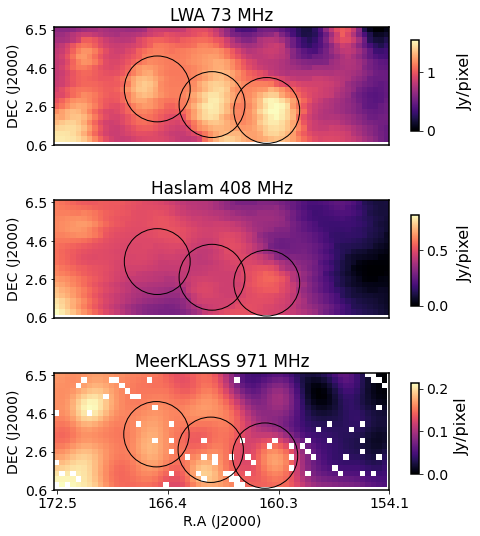}}\\
      \caption{The MeerKLASS pilot observation patch at 1.8$^{\circ}$ resolution as seen at 73, 408 and 971\,MHz. The data are in Jy per pixel and the flux densities above the minimum value in the map at each frequency are shown. The three black circles highlight the regions of investigation.}
         \label{fig:fim}
   \end{figure}

 \subsection{Spectral Energy Distributions}
  \label{sec:apr}
Another useful technique for the combined analysis of numerous different emission maps are spectral energy distribution (SED) plots. Flux density values ($S$) associated with different spatial areas can be used to construct SEDs, and the flux density synchrotron spectral index ($\alpha = \beta + 2$) can then be fitted to the data. We use both the LWA 73\,MHz map and the Haslam 408\,MHz map alongside the MeerKAT data to construct SEDs. For the analysis in this section we choose to use the destriped-only version of the reprocessed Haslam map (`ds' as opposed to `dsds'), as the LWA and MeerKAT data have not had their point source contributions removed. The data were converted from kelvin into Jy per pixel with:
\begin{equation}
\frac{S(p)}{{\rm Jy/px}} = 10^{26}\, \frac{T(p)}{\rm K}  \left( \frac{2\, k_{B} \nu^{2}}{c^{2}} \right) \Omega_{{\rm{pix}}},
\end{equation}
where $ \Omega_{{\rm{pix}}}$ is the pixel solid angle of $0.3 \times 0.3$ square degrees converted into steradians. The maps were then smoothed to a resolution of $1.8^\circ$; \autoref{fig:fim} shows our observation patch as seen by LWA, Haslam, and MeerKAT at this resolution. We would like to plot the evolution of the flux in one pixel over frequency, however to ensure a more robust measurement we instead choose to use the mean temperature within a beam-sized radius. Having both negative and positive values within the radius would skew the mean value therefore, instead of using mean-centred data, the flux density values were shifted so that the minimum value in each map would be equal to zero. Three regions were selected at which to calculate the mean flux density at each frequency. The three circles, each with a radius of $1.8^\circ$, are marked in each of the plots of \autoref{fig:fim}. The MeerKAT data have different flagged pixels at each frequency.

      \begin{figure}
      \centering
            {\includegraphics[width=0.91\linewidth]{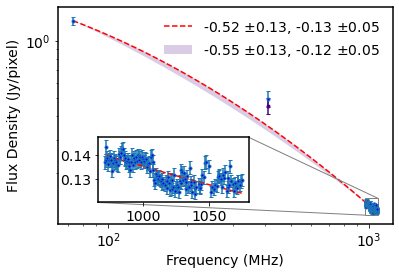}}\\
            {\includegraphics[width=0.91\linewidth]{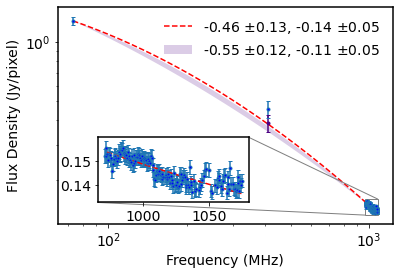}}\\
            {\includegraphics[width=0.91\linewidth]{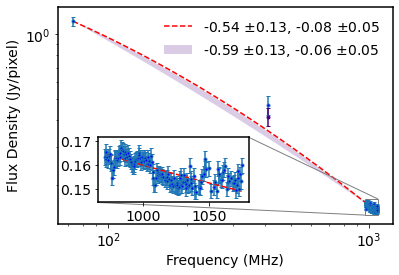}}\\
      \caption{SEDs for $1.8^\circ$ radius regions centred at: {\it{Top:}} (RA, Dec) = (161$^{\circ}$, 2.4$^{\circ}$), {\it{Middle}}: (RA, Dec) = (164$^{\circ}$, 2.7$^{\circ}$), {\it{Bottom}}: (RA, Dec) = (167$^{\circ}$, 3.5$^{\circ}$). Insets show magnified images of the MeerKAT data points. The SEDs are shown for both the fit using Haslam data (red dashed line) and the Commander synchrotron emission map (purple data point and band).}
         \label{fig:apres}
   \end{figure}

SEDs for each of the three highlighted regions are shown in \autoref{fig:apres}. The errors plotted are the standard errors on the mean temperature. The errors used to inform the parametric fit are a combination (in quadrature) of the standard error on the mean and the survey calibration uncertainties. Overlaid onto each plot is a band representing a theoretical synchrotron SED. This band is formed using the Commander diffuse synchrotron emission estimate map \citep{planck_CS_2016} to provide a measurement at 408\,MHz (shown by the purple data point) with a width of 10\% -- the Commander map error budget -- in place of the Haslam measurement. We choose to use the Commander estimate in place of the Haslam measurement to determine the SED as the empirical Haslam data show some disagreement with both the LWA and MeerKAT data, namely a consistent positive offset (as shown in \autoref{fig:apres}). We choose to use the destriped-only version of the reprocessed Haslam map, which gives flux densities of 0.49, 0.47 and 0.47 Jy/pixel for each of the three regions, respectively. If we were to use the source-subtracted and destriped version of the map the flux densities would in fact be lowered to 0.41, 0.44 and 0.45 Jy/pixel. However, as neither the MeerKAT nor the LWA data have had point sources explicitly fitted, removed and inpainted we choose not to use the source-subtracted Haslam data. Overall, the deviations of the destriped-only Haslam data from the fitted spectral forms are smaller than two sigma across all three regions.

The insets for each of the three plots shown in \autoref{fig:apres} reveal the slight `S' shape due to the shape of the receiver bandpasses between 950 and 1075\,MHz. This pattern can be seen in figure 16 of W21, which characterises the mean bandpass for all the MeerKLASS pilot observations. Incidentally, if the MeerKAT data were to be averaged into a single data point representing the temperature at their mean frequency, the fit would then pass through all three data points (one from the LWA, one from Haslam, and one from MeerKAT), and the predicted spectral index curvature would be two to three times higher than the expected value for all three regions. This demonstrates the statistical value of the MeerKAT data -- although the synchrotron emission temperature is weaker at 1\,GHz than at 73\,MHz, the sheer number of measurement channels available is enough to significantly influence the fit.       

We fit the following spectral form to the data:
\begin{equation}
S(\nu) = S(\nu_{0}) \left( \frac{\nu}{\nu_{0}} \right) ^{\alpha (\nu)}, 
\end{equation}
where
\begin{equation} \label{eq:alpha}
\alpha (\nu) \equiv f(\alpha_{0}, c; \nu) = {\alpha_{0} \, + \, c \, {\rm{ln}}(\nu / \nu_{0})},
\end{equation}
and $\nu_{0} = 73$~MHz. Assuming Gaussianity and independence of the data points, we then fit the data to determine $\alpha_{0}$ and $c$, obtaining a posterior probability distribution $p(\alpha_{0}, c)$. To find the mean $\langle \alpha (\nu) \rangle$ and its rms error $\Delta \alpha(\nu)$, we calculate
\begin{align}
\langle \alpha(\nu) \rangle &= \int \int f(\alpha_{0}, c; \nu) p(\alpha_{0}, c) \, d\alpha_{0} \, dc, \\
\langle \alpha^{2}(\nu) \rangle &= \int \int f^{2}(\alpha_{0}, c; \nu) p(\alpha_{0}, c) \, d\alpha_{0} \, dc, \\
\Delta \alpha(\nu) &= \sqrt{\langle\alpha^{2}(\nu)\rangle - \langle\alpha(\nu)\rangle^{2}}, 
\end{align}
where the posterior distribution has been normalised to unity.

\begin{table*}
\centering
\begin{tabular}{||c c c c c c||} 
 \hline
 &{\bf{ARCADE2}} & {\bf{EDGES}} & {\bf{R1}} & {\bf{R2}} & {\bf{R3}} \\
 \hline\hline
 {\bf{Spectral Index}} & $-2.60 -0.081 \, \rm{ln}\frac{\nu}{310}$ & $-2.57 -0.075 \, \rm{ln}\frac{\nu}{75}$  &   $-2.55 -0.12 \, \rm{ln}\frac{\nu}{73}$ & $-2.55 -0.11 \, \rm{ln}\frac{\nu}{73}$ & $-2.59 -0.06 \, \rm{ln}\frac{\nu}{73}$ \\
 {\bf{Galactic Latitude $(^{\circ})$}} & 0 & Full range & 53 & 53 & 53 \\
  {\bf{Resolution FWHM $(^{\circ})$}} & 11.6 & 71.6 & 1.8 & 1.8 & 1.8 \\
 \hline \hline
\end{tabular}
\caption{Synchrotron spectral forms from \citet{kogut} and \citet{edgescurve}, alongside the spectral forms fit from the three SED plots shown in \autoref{fig:apres}.}
\label{tabsed}
\end{table*}

      \begin{figure}
      \centering
            {\includegraphics[width=0.85\linewidth]{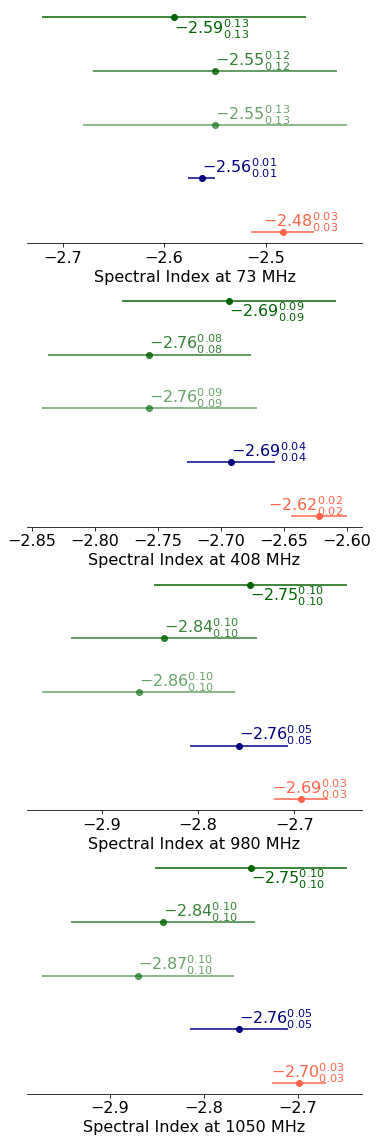}}
      \caption{The predicted synchrotron spectral index at 73, 408, 980 and 1050\,MHz as given by ARCADE2, EDGES and MeerKAT data. The ARCADE2 values are plotted first (closest to the x-axis), followed by EDGES and then the three MeerKAT spectral fits.}
         \label{fig:hubs}
   \end{figure}

The form for the spectral index in Eq.~\ref{eq:alpha} is taken from \citet{kogut}, where the synchrotron spectral index curvature, $c$, is measured between 22 and 1049\,MHz within the Galactic plane ($\lvert b \rvert < 40^{ \, \circ}$) at a resolution of $11.6^\circ$. \citet{kogut} determines the parameter values as $\beta = -2.60 \pm 0.04 $ and $c = -0.081 \pm 0.028$ for $\nu_{0} = 310\,$MHz. In \autoref{fig:apres} this spectral form is fit to each of the three $1.8^\circ$ regions, and the best-fit parameter values are summarised in Table~\ref{tabsed}. We also include the recent results from \cite{edgescurve}, who find $2.59 < \beta < -2.54$ and $-0.11 < c < -0.04$ for $\nu_{0} = 75\,$MHz from the EDGES survey at a resolution of $71.6^\circ$. As EDGES determines this spectral index across the 0 to 12-hour LST range, the full range of Galactic latitudes are probed, but the Galactic longitude of the observation is such that the Galactic centre is avoided.   
   
The predicted $\beta$ values at various frequencies, following each spectral form, are shown in \autoref{fig:hubs}. The four plots are for frequencies 73, 408, 980, and 1050\,MHz. The ARCADE2 values are plotted first (closest to the x-axis), followed by EDGES and then the three MeerKAT spectral fits. Direct comparisons between these results and those of \citet{kogut} cannot be made, as different regions of the sky are under analysis and the synchrotron spectral index is believed to change across the sky -- most notably with Galactic latitude due to particle energy losses \citep{gold}. Strong agreement, to within $1\sigma$, is seen between the MeerKAT data and ARCADE2 and EDGES for all three regions, however.

\section{Comparison with Principal Component Analysis}
\label{sec:PCA}

\begin{figure}
\includegraphics[width=8.1cm]{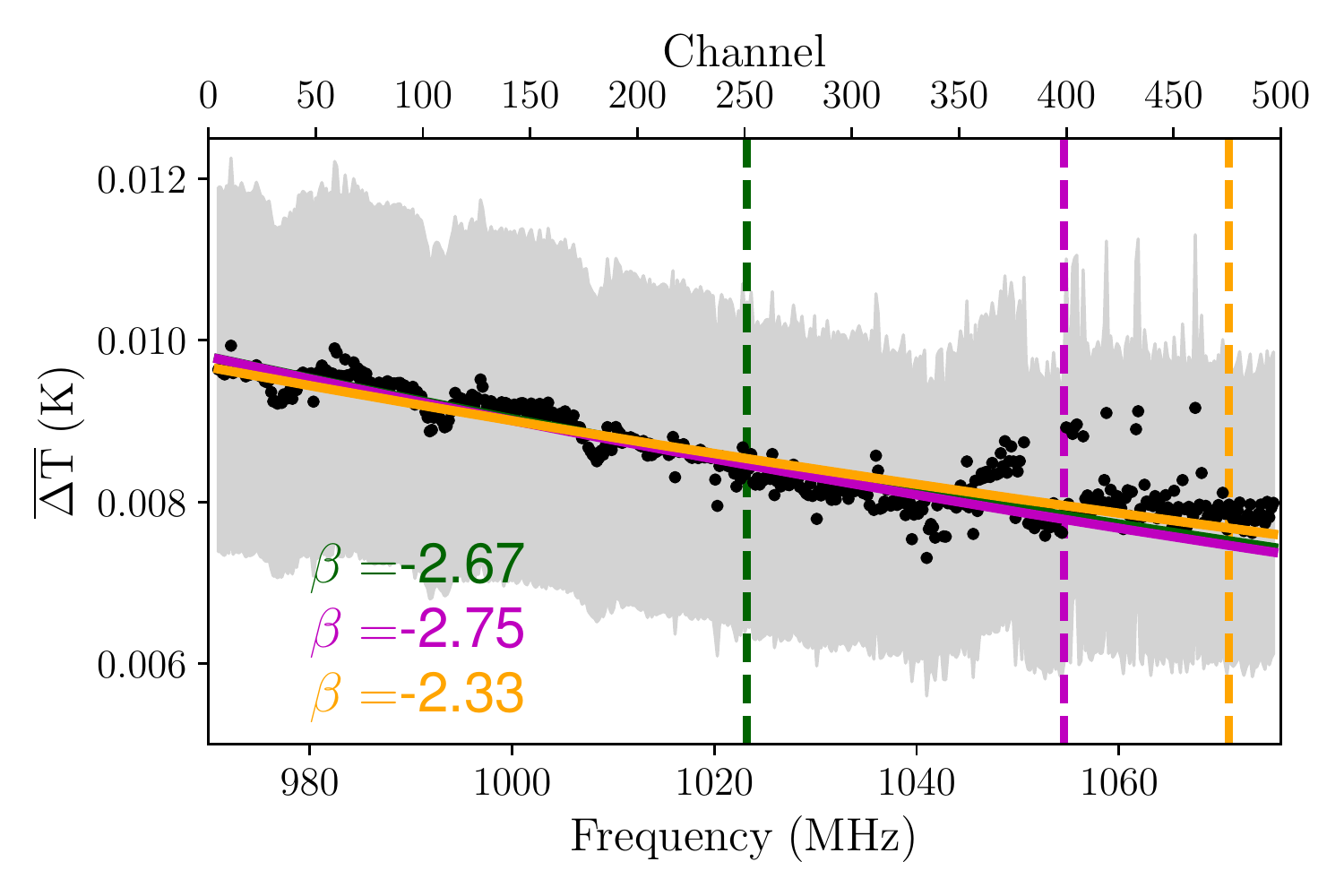}\\
\includegraphics[width=8.1cm]{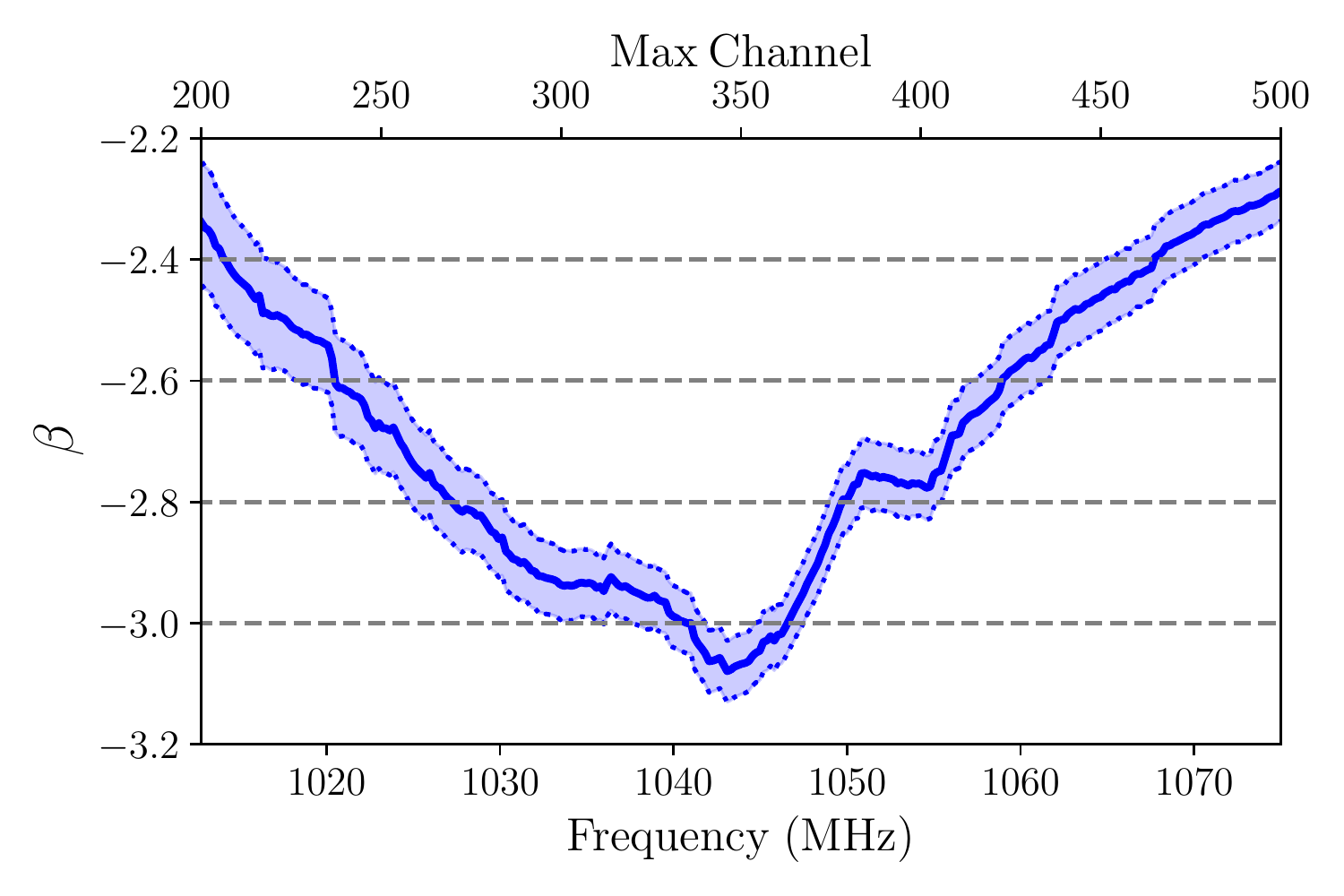}
\caption{\textit{Top:} The mean value (black dots) across pixels of the 1st-component only data-cube as a function of frequency (or channel number) with its $1\sigma$ deviation (in gray). The power-law model resulting from the fit is over-plotted for three different choices of the highest frequency considered, shown as vertical lines with the same colour code ($\sim 1020$~MHz in green, $\sim 1055$~MHz in magenta, and $\sim 1075$~MHz in orange). \textit{Bottom:} The values of the fitted spectral index $\beta$ and their 1$\sigma$ fitting errors as a function of the maximum channel number considered.
}
\label{fig:beta_PCA}
\end{figure}

In this section we complement the results obtained in Section~\ref{sec:res} by analysing the MeerKLASS pilot data maps under a different set of assumptions. As the cosmological 21\,cm signal is orders of magnitude weaker than the astrophysical foregrounds in intensity mapping data, its recovery often relies on Blind Source Separation (BSS) methods for disentangling the different -- astrophysical, cosmological, and spurious -- components of the data cube \citep{blindChallenge}. These methods make general statistical assumptions on the foregrounds, and require no prior knowledge on the 21\,cm signal. BSS can also be used to extract information on the foreground emission itself, as is also done in the CMB context \citep{deOliveira2008, planck_CS_2016, planck_CS_2018}. For intensity mapping at the frequencies of interest, any successful BSS decomposition should isolate the diffuse galactic synchrotron component, as it is the brightest.

Principal Component Analysis (PCA) is the most straightforward BSS algorithm. It assumes that the various components of the measured temperature are uncorrelated. In practice, one computes the frequency-frequency covariance matrix of the data and then orders its eigenvectors by eigenvalues. Foregrounds should inhabit the largest/brightest eigenmodes. Specifically, the first PCA component should be a good proxy for the dominant galactic synchrotron emission. 

We perform a PCA decomposition on the mean-centred maps obtained from combining the three observation blocks described in Section~\ref{sec:mm}, using all available pixels and keeping their native resolutions. We isolate the first component by projecting the first eigenvector out of the original data cube. We cannot take into account the pixels' absolute temperatures, as we started from mean-subtracted maps, however we do expect their behaviour in frequency to follow a power law.

In the top panel of \autoref{fig:beta_PCA} we show the mean value computed across pixels of this first component as a function of frequency, together with the associated $1\sigma$ scatter. \autoref{eq:synch} is used to determine the spectral index assuming $\nu_1 = 1$~GHz. We obtain the best-fit value of $\beta$ using the {\tt curve\_fit} function of \texttt{scipy} that performs a non-linear least-squares fit.  We show the resulting fit for three illustrative values for the number of channels considered. The mean temperature behaviour slightly deviates from a power law at higher frequencies, consistently with the interpretation that the data contain more RFI contamination at these frequencies (see Section~\ref{sec:data}). Although the exact value of the spectral index changes with frequency, the data can always be described with a power law with a reasonable, synchrotron-like value of $\beta$, as can be seen in the lower panel of \autoref{fig:beta_PCA}.
Indeed, $\beta$ evolves from a steeper $-3$ spectral index when including 350 channels to a flatter $-2.3$ spectral index when including all the channels, with a stable fitting error.
The lower panel of \autoref{fig:beta_PCA} reaffirms the need for numerous frequency channels when attempting to determine the spectral index across a small frequency range, as seen in Section~\ref{sec:otf}. 

The blind PCA procedure does not guarantee that the resulting datacube contains only the synchrotron contribution, or even the whole synchrotron contribution, especially as we expect residual systematics to be present in these data. 
With this test, however, we have (1) verified that the BSS process required for the survey's primary science goal works on the pilot data within our expectations; (2) confirmed that we converge on similar spectral indices across the majority of pixels, suggesting that the first principal component can be identified with a physical synchrotron component; and (3) found agreement with the results of Section~\ref{sec:res}, reinforcing the paper's main results.

\section{Conclusions}
 \label{sec:conc}
 
In this work, we have used MeerKLASS pilot data from 64 MeerKAT dishes to study the spectral index of synchrotron emission within $145^{\circ} < \alpha < 180^{\circ}$, $-1^{\circ} < \delta < 8^{\circ}$. The data were taken across 2 months in early 2019 and span 4.5 hours in total. The full MeerKLASS survey will be around 4,000 deg$^{2}$ and is intended as a single-dish 21cm intensity mapping survey. Despite the pilot data only providing a fraction of the planned MeerKLASS sky coverage and observation time, the signal-to-noise ratio has proven sufficient to map the largest 21cm foreground: diffuse Galactic synchrotron emission.
 
We make use of the `T-T plot' method as a way of measuring the spectral index of the dominant emission within a region of the sky without knowing the survey zero-levels of any of the data involved. W21 have already shown that the MeerKAT single-dish combined receiver and elevation-dependant temperature contributions change smoothly across RA/time using this method, and so we adapt the method to operate on chunks of time-ordered data (TOD) small enough for the Galactic and extragalactic emission to be the only time-varying signals. The optimal chunk size, and ultimate suitability of this approach, were determined through tests on simulated data. We then applied our version of the T-T plot method to the MeerKLASS TOD, using the Haslam 408\,MHz data to provide the second observational data set with which to perform a linear regression to recover the synchrotron spectral index.

The TOD were divided into subsets of dishes and the synchrotron spectral index was measured between 408 and 1060\,MHz for each observational block and each subset. Following various cuts on data quality, no significant (greater than 5$\%$) variation of the spectral index was found between the $408-971$\,MHz and $408-1060$\,MHz frequency ranges, nor between observation blocks or dish subsets. Having shown the TOD to be self-consistent, they were then averaged together to produce a map of diffuse synchrotron emission within our observation field. Throughout this work we made the assumption that diffuse free-free emission is negligible at these frequencies and Galactic latitudes.  

We used our estimated map of synchrotron emission alongside data from LWA, Maipu/MU, ARCADE2, and EDGES to probe the synchrotron spectral index and any curvature over frequency that it may have over two orders of magnitude in frequency. Linear regression between the LWA, Maipu/Mu, and MeerKAT data reveals a synchrotron spectral index of $-2.75 < \beta < -2.71$ between 45 and 1055\,MHz, which is consistent with the value of $2.76 \pm 0.11$ found by \citet{plat} between $400 - 7500$\,MHz.

By choosing three compact circular regions, each with $1.8^\circ$ radius, and reconstructing their spectral energy distributions, we fit for the synchrotron spectral index curvature across the LWA, Haslam, and MeerKAT frequency range ($73 - 1075$\,MHz). We find a degree of curvature that changes the spectral index from $-2.55 \pm 0.13$ at 73\,MHz to $-2.87 \pm 0.10 $ at 1050\,MHz, which agrees with the spectral index curvature found by both ARCADE2 and EDGES to within $1\sigma$.

Experiments like the LWA and EDGES have demonstrated that -- although not designed with the primary goal of ISM physics in mind -- 21cm intensity mapping experiments can be well-placed in terms of frequency and angular resolution to contribute to our understanding of diffuse Galactic synchrotron emission. Any improvement to the understanding of our Galaxy also drives improvements in foreground modelling and removal methods, suggesting a symbiotic relationship between ISM physics and cosmology. This was further demonstrated in this work through a comparison between the emission probed using a T-T plot analysis and a Principal Component Analysis. Indeed, some of the analyses performed in service of studying the ISM can also be used to cross-check the quality of foreground cleaning in the intensity mapping data. Applying a PCA decomposition, we found that the first component of our data scales with a power law whose spectral index is in the range $-3 < \beta < -2.3$, depending on the inclusion of the higher frequency data, as expected for the dominant Galactic synchrotron component.

This work also adds to the body of work being built up using interferometers, by providing the complementary single-dish view of synchrotron emission. \citet{jonas} and \citet{gem} have demonstrated the ability to calibrate the zero-level of a radio survey ``on-the-fly'' through linear regression with existing, zero-level calibrated survey data. We apply this technique to highlight how MeerKLASS data can be used alongside existing data to, for instance, add to the knowledge of diffuse Galactic emission represented by the Global Sky Model \citep{gsm}.

To conclude, 21cm intensity mapping experiments are now delving into less-explored MHz frequency ranges. For the first time in radio astronomy we are measuring a wide band of MHz frequencies with small individual channel resolutions, for example the 971.2 to 1075.5\,MHz frequency range with a channel width of 0.2\,MHz studied in this paper. Additionally, MeerKLASS has the excellent signal-to-noise ratio that comes from running 64 single-dish radio astronomy surveys simultaneously, as well as L-band ($900-1670$\,MHz) receivers with remarkably low system temperatures. Therefore, despite observing at high Galactic latitudes, it is possible to measure synchrotron emission and, when combining results across intensity mapping experiments, to probe spectral index curvature, as we have demonstrated. Future observations from MeerKLASS, as well as further data releases from other 21cm intensity mapping experiments, will allow us to extend this work to larger regions of sky.

\section*{Acknowledgements}
This research has made use of the NASA/IPAC Extragalactic Database (NED), which is funded by the National Aeronautics and Space Administration and operated by the California Institute of Technology. We acknowledge use of the following software: {\tt HEALPix} \citep{healpix}, {\tt matplotlib} \citep{matplotlib}, {\tt numpy} \citep{numpy} and {\tt scipy} \citep{2020SciPy-NMeth}.

We acknowledge the use of the Ilifu cloud computing facility, through the Inter-University Institute for Data Intensive Astronomy (IDIA). The MeerKAT telescope is operated by the South African Radio Astronomy Observatory, which is a facility of the National Research Foundation, an agency of the Department of Science and Innovation. We acknowledge support from the South African Radio Astronomy Observatory and National Research Foundation (Grant No. 84156). This result is part of a project that has received funding from the European Research Council (ERC) under the European Union's Horizon 2020 research and innovation programme (Grant agreement No. 948764; PB). PB also acknowledges support from STFC Grant ST/T000341/1. IPC acknowledges support from the `Departments of Excellence 2018-2022' Grant (L.\ 232/2016) awarded by the Italian Ministry of University and Research (\textsc{mur}) and from the `Ministero degli Affari Esteri della Cooperazione Internazionale - Direzione Generale per la Promozione del Sistema Paese Progetto di Grande Rilevanza ZA18GR02'. SC is supported by STFC grant ST/S000437/1. MS acknowledges funding from the INAF PRIN-SKA 2017 project 1.05.01.88.04 (FORECaST).

\section*{Data Availability}
The analysis pipeline software is available on request. Data products are also available on request but unrestricted public access is subject to a 12-month proprietary period starting from the date of this publication. Access to the raw data used in the analysis is public (for access information please contact archive@ska.ac.za).


\bibliographystyle{mnras}
\bibliography{refs} 




\appendix
\section{Selecting regions for T-T plots}\label{sec:apA}

\begin{figure}
\centering
{\includegraphics[width=0.8\linewidth]{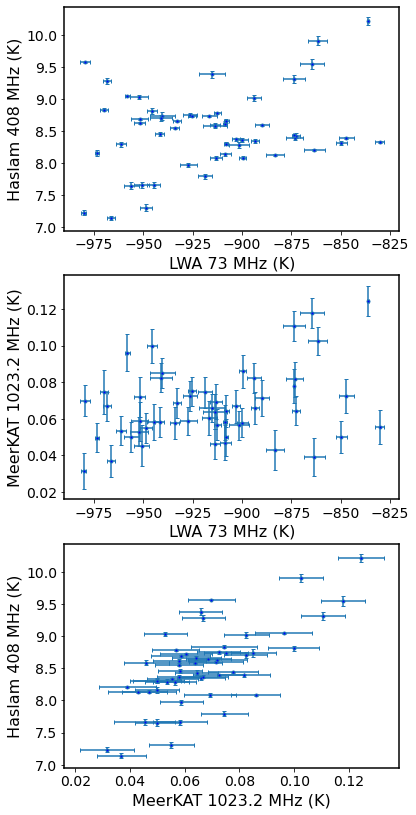}}
\caption{T-T plots between MeerKLASS pilot data at 1023\,MHz, Haslam data at 408\,MHz, and LWA data at 73\,MHz within $163^{\circ} < \alpha < 182^{\circ}$. All data are at 1.8$^{\circ}$ resolution.}
\label{fig:a1}
\end{figure}

Throughout this paper we select data within the $154^{\circ} < \alpha < 163^{\circ}$ region to perform our T-T plot analysis. \autoref{fig:a1} shows T-T plots between the Haslam, LWA and 1023\,MHz MeerKAT data for the $163^{\circ} < \alpha < 182^{\circ}$, region. It can be seen that within this higher RA region there is insufficient spatial structure above the survey noise levels with which to perform a linear regression between data sets. Therefore we choose to exclude this region of RA from all of our T-T plots.

\section{Verification on simulation data}
\label{sec:apB}

\begin{figure}
\centering
{\includegraphics[width=0.9\linewidth]{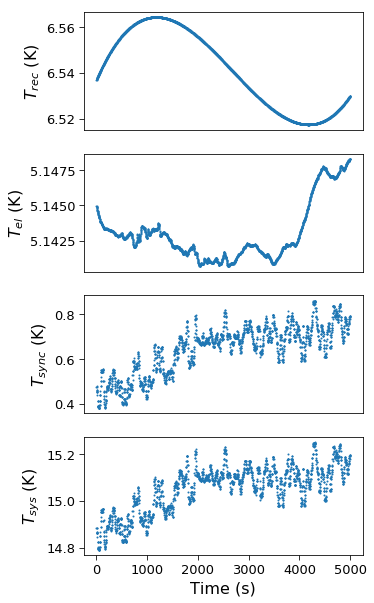}}\\
\caption{Models for the receiver, combined atmospheric and ground pick-up, synchrotron emission and total system temperature for a typical MeerKAT single dish at a single frequency.}
 \label{fig:cosims}
  \end{figure}

As the MeerKAT data are more complex than simply synchrotron emission plus a constant additive temperature, we first apply the T-T plot method to simulated data to test its robustness in the face of terrestrial, spatially-varying temperature components. Thanks to the instrumental characterisation performed in W21, realistic models for the receiver temperature, atmospheric, and ground pick-up emission components are readily available. \autoref{fig:cosims} shows the model receiver temperature and the combined atmospheric and ground pick-up temperature for a single MeerKAT dish at a single frequency as a function of time as it scans the survey area. To produce the total system temperature model, the CMB monopole temperature was added alongside a model for the diffuse synchrotron emission. The synchrotron emission model is the Haslam 408\,MHz data smoothed to $1.6^\circ$ resolution and extrapolated up to the MeerKAT frequencies using a spectral index which varies randomly between pixels according to a Gaussian distribution centred at $-2.9$ with a standard deviation of 0.03. The value of $-2.9$ was chosen as it is the mean value of the synchrotron spectral index map between 0.408 and 23\,GHz presented in \citet{mamd08}. The receiver temperature can be seen to vary smoothly over time with an overall change in temperature of around 0.04\,K, while the combined atmospheric and ground-pick up emission is more stable over time with an overall temperature change of around 0.01\,K. The smallest component in terms of its median temperature is the diffuse synchrotron emission; this emission is seen to vary with position on the sky which, due to our scan strategy, results in what appears to be rapid variations over time when compared to the terrestrial temperature contributions. This component dominates the 0.4\,K temperature change over time seen in the system temperature.

   \begin{figure}
   \centering
 {\includegraphics[width=0.9\linewidth]{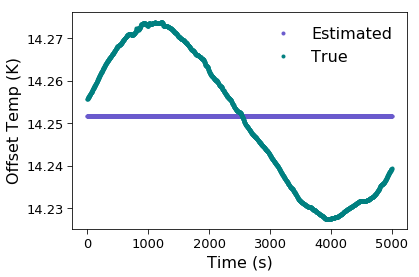}}\\
  {\includegraphics[width=0.9\linewidth]{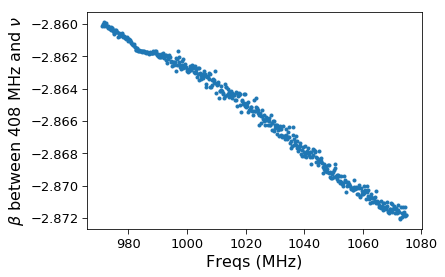}}\\
 \caption{{\it{Top panel:}} The actual offset temperature within one frequency channel of our simulated total system temperature across time, alongside the value estimated by dividing the data into 30s time chunks. {\it{Bottom panel:}} The spectral index between the simulated data representing MeerKAT channels below 1075\,MHz and the Haslam 408 MHz data within the MeerKLASS pilot observation patch.}
 \label{fig:offsims}
   \end{figure}

Performing a linear regression between Haslam data and the full hour and ten minutes worth of MeerKAT observational data at a given frequency will not provide an accurate estimate of the synchrotron spectral index because the time-varying terrestrial contributions to the total system temperature violate the main principal of the T-T plot method -- that diffuse Galactic emission is the only time-varying component in the data. To combat this, the data are divided into smaller time chunks in order to isolate time periods in which the only significantly time-varying component is the diffuse Galactic emission. These smaller time chunks will naturally contain fewer data points than the full observation block, and so the fitted gradient from the linear regression will be noisier. We expect only small deviations from the mean spectral index across the $154^{\circ} < \alpha < 163^{\circ}$ region however. Therefore, we split the data into time chunks, perform the linear regression to determine the spectral index and temperature offset for each time chunk, and then use the weighted mean of these values to provide one single temperature offset and spectral index value at each frequency. The optimum chunk size of 30\,s was determined empirically from these simulations, and is balance between finding a time short enough for the receiver temperature to be considered constant and long enough to observe enough data to get good fits.
   
The top panel of \autoref{fig:offsims} shows the true offset (CMB monopole plus receiver temperature plus the combined atmospheric and ground-pick up contributions) for our simulated data at one frequency as a function of time, alongside our estimated value. As the estimated offset is just the mean offset from all the 30\,s time chunks, it is a single value and cannot capture the smooth variation in temperature over time. From the top panel of \autoref{fig:offsims} we can see that the estimated offset differs from the true offset by a maximum of 0.02\,K. The $\sim 0.02$\,K accuracy of this estimate is satisfactory when compared to the $\sim 0.4$\,K variations in Galactic synchrotron emission temperature. The low number of data points available in each 30\,s time chunk results in individual offset and gradient estimates that are too noisy to trust individually, but the weighted mean of these values provides good estimates for both the temperature offset and emission spectral index. The bottom panel of \autoref{fig:offsims} plots the fitted, mean (across pixels), spectral index between Haslam data at 408 MHz and simulated MeerKAT data as a function of frequency. The time chunk implementation of the T-T plot method can be seen to recover the correct mean spectral index, which is a constant across the full frequency range in these simulations, to within $0.04$ of the input value of $-2.9$.

\balance


\bsp	
\label{lastpage}
\end{document}